\documentclass[11pt,a4paper]{article}

\usepackage{amsmath}
\usepackage{amssymb}
\usepackage{cite}
\usepackage{graphicx}
\usepackage{bbm}

\usepackage[latin1]{inputenc}
\usepackage{tikz}
\usetikzlibrary{shapes,arrows}

\newcommand{\be}{\begin{equation}}  
\newcommand{\ee}{\end{equation}}

\begin{document}

\begin{titlepage}

\begin{flushright}

\end{flushright}

\vskip 1.35cm
\begin{center}
{\bf\Large 
~~~~~Higher Order BLG Supersymmetry Transformations 
\newline

from 10-Dimensional Super Yang Mills 
}

\vskip 1.2cm
John Hall$^{a}$\footnote{email John.Hall100@googlemail.com}, ~~~Andrew Low$^{b}$\footnote{email Andrew.Low@wim.gdst.net}
\vskip 0.4cm
{\it $^{a}$ Alumnus of Imperial College, South Kensington, London \\
$^{b}$ Physics Department, Wimbledon High School, Mansel Road, London, SW19 4AB}
\vskip 1.5cm

\abstract{\noindent
We study a Simple Route for constructing the higher order Bagger-Lambert-Gustavsson theory - both supersymmetry transformations and Lagrangian - starting from knowledge of only the $10$-dimensional Super Yang Mills Fermion Supersymmetry transformation. We are able to uniquely determine the four-derivative order corrected supersymmetry transformations, to lowest non-trivial order in Fermions, for the most general three-algebra theory. For the special case of Euclidean three-algbera, we reproduce the result presented in arXiv:$1207.1208$, with significantly less labour. In addition, we apply our method to calculate the quadratic fermion terms in the higher order BLG fermion supersymmetry transformation.

}
\end{center}
\end{titlepage}

\newpage
\tableofcontents
\newpage

\setcounter{page}{3}

\section{Introduction}
This short article is about formulating a Simple Route from the 10-dimensional Super Yang Mills (SYM) Fermion Supersymmetry Transformation to the full $(2+1)$ dimensional Bagger-Lambert-Gustavsson (BLG) theory. 

The BLG Lagrangian and supersymmetry transformations \cite{Bagger:2007st, Gustavsson:2007bk} can be thought of as the leading order terms in an $l_p$ expansion of a non-linear M2-brane theory. This is analogous to how Super Yang Mills theory represents the leading order terms of the Born-Infeld action, which describes the dynamics of coincident D-branes. In Bagger and Lambert's original paper, dimensional analysis was used alongside a novel algebraic structure to write down the most general scalar, fermion and gauge field supersymmetry transformations. The supersymmetry algebra was shown to close on to equations of motion which were used to infer the structure of the Lagrangian. In \cite{Richmond1}, Richmond used a similar approach to determine the next-to-leading order four-derivative corrected supersymmetry transformations and Lagrangian of the Euclidean BLG theory. Starting from the most general expressions allowed by dimensional analysis, he was able to uniquely determine the coefficients through the invariance of the Lagrangian and closure of the supersymmetry algebra. 

An alternative approach for determining the Lorentzian BLG Lagrangian, at lower and higher order, was presented in a series of papers \cite{Mukhi:2008hy},\cite{Mukhi:2008qx},\cite{Mukhi:2008er} in which the authors used a duality transformation due to de-Witt, Nicholai and Samtleben (dNS) \cite{Nicolai, deWit1, deWit2}. The duality is based on the idea that a gauge field is dual to a scalar in $(2+1)$ dimensions and it is therefore possible to replace the gauge field with a scalar and in so doing enhance the $SO(7)$ symmetry of the scalars to $SO(8)$. In \cite{Mukhi:2008er}, this approach was applied to the $\alpha^{'2}$ terms of the D2-brane Lagrangian in order to determine the four-derivative corrections to the Lorentzian BLG theory. Furthermore, it was shown that all higher-order terms were expressible in terms of three-brackets $[X^I, X^J, X^K]$. This led the authors to conjecture that the higher-order Lagrangian they had derived would also apply to the Euclidean BLG theory. This conjecture was confirmed in \cite{Mukhi:2009ra} where the authors used dimensional analysis to write down all possible terms at four-derivative order, and then applied the Novel Higgs mechanism to match coefficients with terms in the D2-brane Lagrangian. This confirmed that the structure of the Lorentzian theory derived using dNS duality had exactly the same form as the Euclidean BLG theory at four-derivative order.

Motivated by this approach, one might consider applying the dNS duality transformation directly at the level of supersymmetry transformations. In \cite{Low1} the four-derivative corrected BLG fermion supersymmetry transformation was derived by applying the dNS duality to the $\alpha^{'2}$ corrections of the non-abelian D2-brane theory. However, the dNS duality was shown to break down when applied to the D2-brane gauge field and scalar field supersymmetry transformations. For a more detailed discussion see \cite{Low2}.  

In this paper we propose a new and simple route for determining the four-derivative corrected BLG supersymmetry transformations, including quadratic fermion terms. Starting from the $\alpha^{'2}$ fermion supersymmetry transformation of ten-dimensional SYM theory, we reduce to $(2+1)$ dimensions and apply the dNS duality to the D2-brane supersymmetry transformation. The resulting SO(8) invariant BLG fermion transformation is used to construct the supercharge, which in turn is used to generate the scalar field and gauge field supersymmetry transformations. The requirement that the supercharge should generate the gauge field supersymmetry transformation constrains the Poisson-bracket structure for the spatial components of the gauge field.

The structure of this article is as follows. In Section 2, we outline our methodology and apply it to the lowest order ten-dimensional SYM fermion transformation to derive the lowest order BLG theory. In Section 3, we apply our method to determine the four-derivative corrected BLG supersymmetry transformations. In Section 4, we apply our method to determine, for the first time, the quadratic fermion terms in the higher order fermion supersymmetry transformations of BLG theory. The Appendix outlines conventions, useful identities and key formulae used in this paper.

\tikzstyle{block} = [rectangle, draw, fill=blue!20, 
   text width=20em, text centered, rounded corners, minimum height=2.5em]
\tikzstyle{line} = [draw, -latex']
\tikzstyle{cloud} = [draw, ellipse,fill=red!20, node distance=1.6cm,  text width=20em,
    minimum height=2.5em]
  
\begin{figure}
\vspace*{-100pt}\caption{The Simple Route to the BLG theory - both Susy transformations and Lagrangian - starting from knowledge of only the 10d Super Yang Mills Fermion Susy transformations.}
 
\vspace*{+20pt}\centering
\begin{tikzpicture}[node distance = 1.6cm,align=flush center]
    \node [block] (SYM) {\small{Start with 10 dim SYM Susy Fermion $\delta \psi $}};
    \node [cloud, below of=SYM] (dimred) {\small{dimensionally reduce to 2+1 dim}};
    \node [block, below of=dimred] (D2) {\small{Non-abelian D2 Brane Susy Fermion $\delta \psi $}};
    \node [cloud, below of=D2] (dualise) {\small{dualise YM gauge field $A^{M}$ to scalar}};
    \node [block, below of=dualise] (BLFermion) {\small{2+1 dim BLG Susy Fermion $\delta \psi $ }};
    \node [cloud, below of=BLFermion] (Supercurrent) {\small{derive supercurrent, generate scalar, gauge field}};
    \node [block, below of=Supercurrent] (Susytrans) {\small{2+1 dim BLG Susy Fermion, Scalar, Gauge Field $\delta \psi,\delta X^{I},\delta A^{\mu }$}};
    \node [cloud, dashed, fill=red!5, below of=Susytrans] (Algebra) {\small{close the algebra for fermion, gauge field}};   
    \node [block, dashed, fill=blue!5, below of=Algebra] (EOM) {\small{Fermion, Gauge Field E.O.M.}};
    \node [cloud, dashed, fill=red!5, below of=EOM] (Supervn) {\small{calculate supervariation of fermion e.o.m.}};
    \node [block, dashed, fill=blue!5, below of=Supervn] (Scalareom) {\small{Scalar Field Equation of Motion }};
    \node [cloud, dashed, fill=red!5,below of=Scalareom] (Integ) {\small{integrate e.o.m.}}; 
    \node [block, dashed, fill=blue!5, below of=Integ] (Lagr) {\small{BLG Lagrangian and it's Susy transformations}};
    \path [line,dashed] (SYM) -- (dimred);
    \path [line,dashed] (dimred) -- (D2);
    \path [line,dashed] (D2) -- (dualise);
    \path [line,dashed] (dualise) -- (BLFermion);
    \path [line,dashed] (BLFermion) -- (Supercurrent);
    \path [line,dashed] (Supercurrent) -- (Susytrans);
    \path [line,dashed] (Susytrans) -- (Algebra);  
    \path [line,dashed] (Algebra) -- (EOM);
    \path [line,dashed] (EOM) -- (Supervn);
    \path [line,dashed] (Supervn) -- (Scalareom);
    \path [line,dashed] (Scalareom) -- (Integ);    
    \path [line,dashed] (Integ) -- (Lagr);   
\end{tikzpicture}
\end{figure}
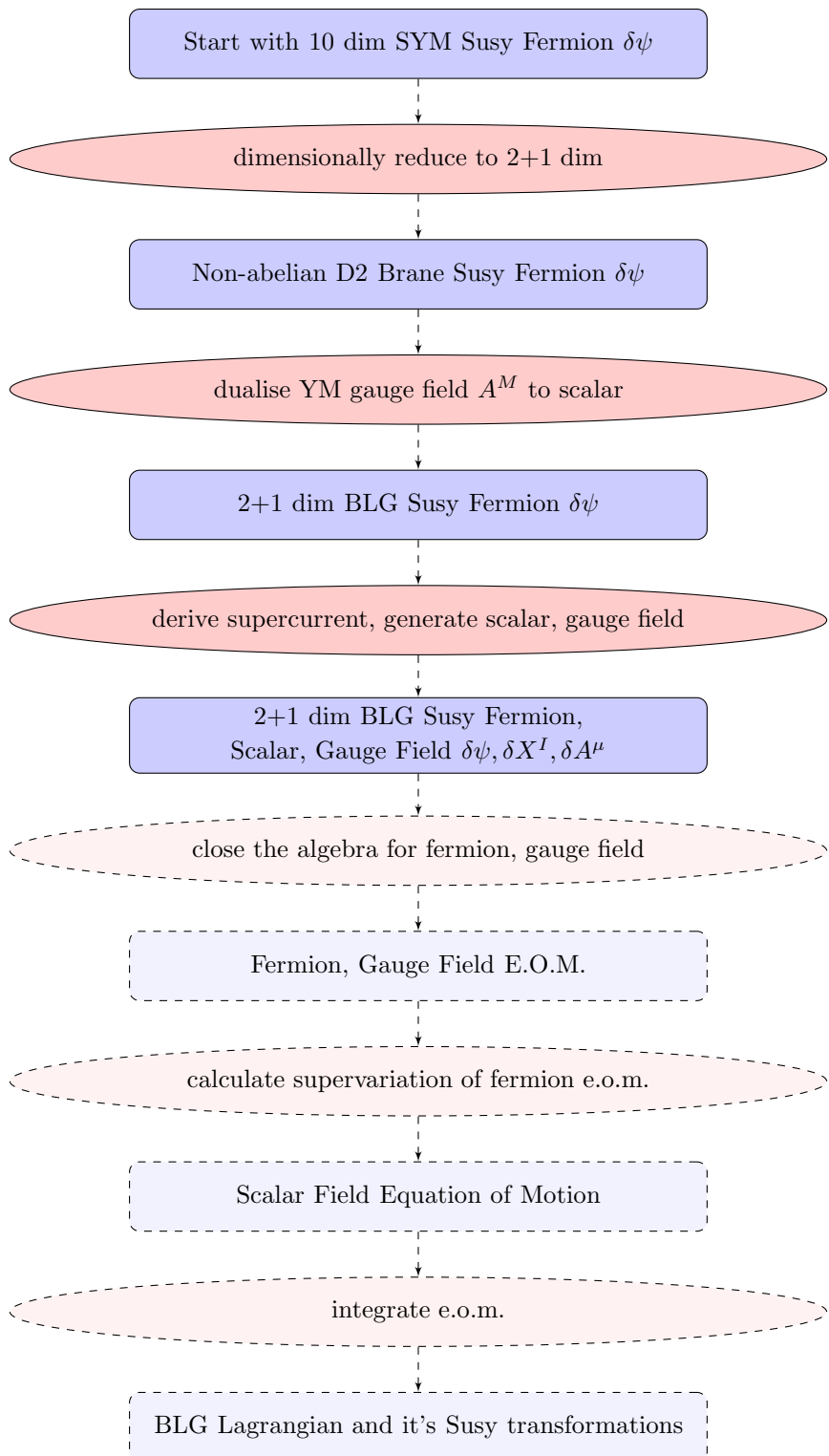

\section{The Simple Route}

Our start point is the 10-dimensional Super Yang Mills Fermion supersymmetry transformation. In 10 dimensions a gauge field has mass dimension $[A] = 4$ whereas a fermion field has mass dimension $[\psi] = 4\frac{1}{2}$. Furthermore, the supersymmetry parameter has mass dimension $[\epsilon] =-\frac{1}{2}$ which follows from the simple observation that the supervariation of a fermionic field must give a bosonic field. A little thought reveals that the fermion transformation of 10d SYM must take the form $\delta \psi = \delta \psi_{(1)} + \delta \psi_{(2)} + \delta \psi_{(3)}$
\begin{eqnarray}\label{aab}
\delta \psi_{(1)} &=&  \; \frac{1}{2} \Gamma ^{MN} F_{MN} \epsilon \\
 \delta \psi_{(2)} & = &  \; \alpha ^{'2} (\lambda _{1} \Gamma ^{MN} F_{PQ} F^{PQ} F_{MN}\epsilon \nonumber\\ 
& & {} +  \; \lambda _{2} \Gamma ^{MN} F_{MP} F^{PQ} F_{QN}\epsilon \nonumber\\
 & & {}  +  \; \lambda _{3} \Gamma ^{MNPQRS} F_{MN} F_{PQ} F_{RS}\epsilon) \label{sec}\\ 
\delta  \psi_{(3)}  \;&=&  \; \alpha ^{'2}( \lambda_4 \bar{\psi} \Gamma^M D^N \psi F_{MN} \epsilon   +    \lambda_5 \bar{\psi } \Gamma ^{MNP} D_{M}\psi  F_{NP} \epsilon) 
\end{eqnarray}
where $M,N$ are the 10 dimensional Lorentz indices taking values $ (0,1,2 \ldots 9)$, $F^{MN}$ is the non-Abelian gauge field strength and $\psi $ represents a ten dimensional complex Majorana-Weyl spinor. The Gamma matrices satisfy the 10-dimensional Clifford algebra. Furthermore we see that the fermion transformation is comprised of three parts: $\delta \psi_{(1)}$ represents the known lowest order fermion supersymmetry transformation, $\delta \psi_{(2)}$ represents the trivial $\alpha^{'2}$ correction and $\delta \psi_{(3)}$ represents a quadratic fermion $\alpha^{'2}$ correction.\footnote{The structure of the $\alpha^{'2}$ corrections to 10d SYM was first investigated by Berghoeff and collaborators in \cite{Bergshoeff:1987uv} \cite{Bergshoeff:2001uvv}.}  The spinors appearing in 10 dimensional Super Yang Mills are Majorana-Weyl and satisfy
\begin{equation}
\Gamma^{(10)} \Psi = \Psi
\end{equation}
where $\Gamma^{(10)}$ is the ten dimensional chirality matrix. Since we are ultimately interested in applying a duality transformation to lift the D2-brane supersymmetry transformations to M-theory it is desirable to look for an embedding of $SO(1,9)$ into $SO(1,10)$ in which $\Gamma^{(10)}$ becomes the eleventh gamma matrix. We denote the gamma matrices of $SO(1,10)$ as $\Gamma^{M} (M = 0, \ldots, 9, 10)$. In eleven dimensions the spinors will be Majorana. The presence of the M2 brane breaks the Lorentz symmetry as $SO(1,10) \rightarrow SO(1,2) \times SO(8)$ and therefore we can have a Weyl spinor of $SO(8)$. Let us denote the chirality matrix of $SO(8)$ by $\Gamma$ where
\begin{equation}
\Gamma = \Gamma^{3 \ldots 9 (10)}
\end{equation}
Half of the supersymmetry of the vacuum is broken by the presence of the M2-brane. We choose conventions in which
\begin{equation}
\Gamma \epsilon = \epsilon, \qquad \Gamma \psi = -\psi
\end{equation}
Under dimensional reduction, the $(9+1)$ dimensional gauge field will split into a $(2+1)$ dimensional gauge field $A_\mu$ and a scalar field $X^i$ transforming under $ SO(7)$. As is usual with dimensional reduction, the fields are independent of the compact directions and therefore one can set $\partial_i = 0$. 


As an illustrative exercise, we derive the lowest order BLG supersymmetry transformations. The only formula required is the lower order 10d SYM fermion transformation
\begin{eqnarray} \label{aah}
 \delta \psi &=&\frac{1}{2} \Gamma ^{MN}F_{MN}\epsilon 
\end{eqnarray}
In what follows we will label $\mu = 0,1,2$ and $i = 1, \ldots , 7$ with the ten dimensional chirality matrix relabelled as $\Gamma^{(10)} = \Gamma^8$. Dimensional reduction of \eqref{aah} results in the following $(2+1)$ dimensional expression
\begin{eqnarray}\label{aai}
 \delta \psi &=& \frac{1}{2} \Gamma ^{\mu \nu }F_{\mu \nu }\epsilon +\Gamma ^{\mu} \Gamma^i D_{\mu} X_i \epsilon -\frac{1}{2}\Gamma ^{ij }X_{ij }
\end{eqnarray}
where $X^{ij} = [X^i, X^j]$. We now consider the effect of applying dNS duality at the level of supersymmetry transformations.\footnote{for a detailed discussion of dNS duality and its relation to BLG theory see for example \cite{Mukhi:2008qx}, \cite{Mukhi:2008er}} This simply involves making the replacement $F_{\mu \nu} = - \epsilon_{\mu \nu \lambda} D^\lambda X^8$. Performing the duality transformation on the fermion transformation leads to
\begin{eqnarray}\label{aabj}
 \delta \psi &=& \frac{1}{2} \Gamma ^{\mu \nu }( -\epsilon _{\mu \nu \lambda} D^{\lambda }X^{8} )\epsilon +\Gamma ^{\mu} \Gamma ^{i}D_{\mu }X^{i}\epsilon +\frac{1}{2} \Gamma ^{ij }(-X_{ij})\epsilon \nonumber \\
 &=& +\Gamma ^{\mu} \Gamma^{I} D_{\mu}X^{I} \epsilon  -\frac{1}{6} \Gamma ^{IJK}X_{IJK}\epsilon 
\end{eqnarray}
which matches the known BLG result. The next step in the process is to use the BLG fermion transformation to derive an expression for the supercurrent. The conserved supercurrent is the Noether current associated with global supersymmetry transformations. Noethers theorem asserts that corresponding to every global symmetry there exists a corresponding conserved current. The usual approach for constructing such an expression is to check the invariance of the Lagrangian under supersymmetry transformations. As is well known, the Lagrangian need only be invariant up to a total derivative to ensure that the Action is invariant. Importantly, the total derivative contributes to the the conserved Noether current. However, we are assuming that we have no knowledge of the Lagrangian and therefore must use an alternative approach for determining the structure of the conserved supercurrent. In \cite{Low3, Low4} it was noted that the supercurrent corresponding to lowest order BLG theory could be derived through knowledge of only the BLG  fermion transformation, in particular  
\begin{eqnarray}\label{aad}
\bar{\epsilon} J^{\sigma } &=&-\bar{\psi } \Gamma ^{\sigma } \delta \psi. 
\end{eqnarray} 
Importantly we emphasise that this expression only requires knowledge of the fermion supersymmetry transformation. In the case of lowest order BLG, constructing the supercurrent results in the following expression
\begin{eqnarray}\label{aal}
+\bar{\epsilon} J^{\sigma} =-\bar{\psi } \Gamma ^{\sigma } \delta \psi  = -(\bar{\psi } \Gamma ^{\sigma }\Gamma ^{\mu} \Gamma^{I} D_{\mu}X^{I} \epsilon)  +\frac{1}{6}(\bar{\psi } \Gamma ^{\sigma } \Gamma ^{IJK}X_{IJK}\epsilon).  
\end{eqnarray}
The validity of this expression can be tested by observing whether the corresponding supercharge generates the expected supersymmetry transformations. The supercharge is the integral over the spatial worldvolume coordinates of the timelike component of the supercurrent
\begin{eqnarray}
Q &=& \int d^2 \sigma J^0  \\ \nonumber
&=& - \int d^2 \sigma \left( D_\nu X^I \Gamma^\nu \Gamma^I \Gamma^0 \psi + \frac{1}{6} X^{IJK} \Gamma^{IJK} \Gamma^0 \psi \right).
\end{eqnarray}
Since the supercharge is the generator of supersymmetry transformations it should be possible to generate the scalar field and gauge field supersymmetry transformations explicitly. 
\subsection*{Scalar transformation}
Let us now use the expression for the supercharge to generate the scalar field supersymmetry transformation
\begin{eqnarray}\label{aae}
\delta X^{I} &=& i\bar{\epsilon }\; [Q,X^{I}] \\ \nonumber
&=& i \bar{\epsilon} [ - \int d^2 \sigma  \left( \partial_\nu X^J (\sigma) \Gamma^\nu \Gamma^J \Gamma^0 \psi (\sigma)\right) , X^I (\sigma' ) ] \\ \nonumber
&=& - i \bar{\epsilon} \Gamma^0 \Gamma^J \Gamma^0 \psi (\sigma)\int d^2 \sigma [\partial_0 X^J (\sigma) , X^I (\sigma'   )] \\ \nonumber
&=& i \bar{\epsilon} \Gamma^J \psi (\sigma)\int d^2 \sigma \delta^{IJ} \delta (\sigma - \sigma'  ) \\ \nonumber
&=& i \bar{\epsilon} \Gamma^I \psi
\end{eqnarray} 
which is the expected form of the BLG scalar supersymmetry transformation.
\subsection*{Gauge Field transformation}
In contrast to the scalar field, the gauge field Poisson Bracket is ill-defined since the gauge field is non-dynamical in $(2+1)$ dimensions. The problem can be traced to the fact that there is no momentum conjugate to the gauge field; this can be seen explicitly at the level of the Lagrangian by looking at the structure of the Chern-Simons term. Motivated by the requirement that the supercharge should generate the gauge field supersymmetry transformation, we have found that by assuming\footnote{we would like to thank Professor Neil Lambert for enlightening discussions surrounding this issue.}
\begin{eqnarray}\label{aag}
[A_{i},A_{j}]_{P.B}=\epsilon_{ij}
\end{eqnarray}
we are able to generate the correct expression for the \emph{spatial} components of the gauge field supersymmetry transformation.
We then conjecture that the same structure holds for the time-like component of the gauge field transformation. For example, at lower order we have
\begin{eqnarray}\label{aap}
\delta A_j&=&i\bar{\epsilon }\; [Q,A_j]_{P.B}\nonumber\\
&=& -(i\bar{\epsilon }  \Gamma ^{i }\Gamma ^{0 } \Gamma ^{I}\psi  ) \int \; d^{2} \sigma \;[A_i ,A_j ]_{P.B} X^{I} \\ \nonumber
  &=& -(i\bar{\epsilon }  \Gamma ^{i}\Gamma ^{0 } \Gamma ^{I}\psi  ) \epsilon_{ij} X^{I}\nonumber \\
  &=& -(i\bar{\epsilon }  \Gamma ^{12}\Gamma ^{i } \Gamma ^{I}\psi  ) \epsilon_{ij} X^{I}\nonumber\\
  &=& +i\bar{\epsilon }  \Gamma _{j} \Gamma ^{I}X^{I}\psi 
\end{eqnarray}
which is the correct expression for the spatial components of the BLG gauge field supersymmetry transformation. We then conjecture that the expression for the spatial components can be generalised to all world-volume indices 
\begin{eqnarray}\label{aar}
\delta A_{\mu  }  &=& +i\bar{\epsilon }  \Gamma _{\mu } \Gamma ^{I}X^{I}\psi. 
\end{eqnarray}
In the next section we will show that our method can be successfully applied to determine the higher-order corrections to the BLG supersymmetry transformations.

\section{Higher Order BLG Supersymmetry Transformations}
In this section we will apply our method to the $\delta \psi_{(2)}$ terms appearing in \eqref{sec} in order to uniquely determine the  higher order BLG supersymmetry transformations. We will then show that for the case of Euclidean BLG theory, our results match the literature \cite{Richmond1}. Our start point is the ten dimensional $\alpha^{'2}$ corrected SYM Fermion transformation
\begin{eqnarray}\label{aat}
 \delta \psi_{(2)} & = &  \; \alpha ^{'2} (\lambda _{1} \Gamma ^{MN} F_{PQ} F^{PQ} F_{MN}\epsilon \nonumber\\ & & {} +  \; \lambda _{2} \Gamma ^{MN} F_{MP} F^{PQ} F_{QN}\epsilon\nonumber\\
 & & {}  +  \; \lambda _{3} \Gamma ^{MNPQRS} F_{MN} F_{PQ} F_{RS}\epsilon) \qquad  
\end{eqnarray}
The first step is to reduce this expression to $(2+1)$ dimensions and then apply the dNS duality transformation. The requirement of $SO(8)$ invariance places constraints on the coefficents appearing in \eqref{aat}. The full derivation of the higher order BLG Fermion transformation can be found in \cite{Low1} and therefore we will only include an illustrative example of how the coefficients can be fixed by looking at an `abelian' truncation of the full theory. In this case, dimensional reduction of \eqref{aat} results in
\begin{eqnarray}\label{aau}
 \delta \psi & = & +\lambda _{1}(\Gamma ^{\mu \nu }F_{\mu \nu }F^{\rho \sigma }F_{\rho\sigma  } +2\Gamma ^{\mu \nu }F_{\mu \nu }\partial^{\rho  }X^{i}\partial_{\rho }X^{i} \nonumber\\
& & { }+2\Gamma ^{\mu }\Gamma ^{i}\partial_{\mu }X^{i}F^{\rho \sigma }F_{\rho\sigma  }+4\Gamma ^{\mu }\Gamma ^{i}\partial_{\mu }X^{i}\partial^{\nu }X^{j}\partial_{\nu }X^{j})\epsilon  \nonumber\\
& & { }  +\lambda _{2}(\Gamma ^{\mu \nu }F_{\mu \rho  }F^{\rho \sigma }F_{\sigma \nu  } -2\Gamma ^{\mu \nu }F_{\mu \rho  }\partial^{\rho  }X^{i}\partial_{\nu  }X^{i}  \nonumber\\
& & { }+2\Gamma ^{\mu }\Gamma ^{i}F_{\mu \rho }F^{\rho\sigma  }\partial_{\sigma  }X^{i}-2\Gamma ^{\mu }\Gamma ^{i}\partial_{\mu }X^{j}\partial^{\rho  }X^{j}\partial_{\rho  }X^{i})  \nonumber\\
& & { } -\Gamma ^{ij}\partial_{\rho  }X^{i} F^{\rho \sigma  } \partial_{\sigma   }X^{j})\epsilon  \nonumber\\
& & { }  + \lambda _{3}(-8 \Gamma ^{\mu \nu \rho }\Gamma ^{ijk}\partial_{\mu    }X^{i}\partial_{\nu    }X^{j}\partial_{\rho    }X^{k})\epsilon.
\end{eqnarray}
Duality is implemented at higher order by making the replacement $   F_{\mu \nu }=+  \epsilon _{\mu \nu \lambda }D^{\lambda } X^{8}$. We are considering the `abelian' truncation in which case the covariant derivatives will be replaced by partical derivatives. A small amount of algebra leads to the following expression
\begin{eqnarray}\label{aav}
 \delta \psi & = & \lambda _{1}(+4\Gamma ^{\mu }\partial_{\mu }X^{8}\partial^{\nu }X^{8}\partial_{\nu }X^{8} -4\Gamma ^{\mu }\partial_{\mu }X^{8}\partial^{\rho  }X^{i}\partial_{\rho }X^{i}  \nonumber\\
& & { }-4\Gamma ^{\mu }\Gamma ^{i}\partial_{\mu }X^{i}\partial_{\nu }X^{8}\partial^{\nu }X^{8}+4\Gamma ^{\mu }\Gamma ^{i}\partial_{\mu }X^{i}\partial^{\nu }X^{j}\partial_{\nu }X^{j})\epsilon  \nonumber\\
& & { }  +\lambda _{2}(-2\Gamma ^{\mu }\partial_{\nu }X^{8}\partial^{\nu }X^{8}\partial_{\mu }X^{8}-2\Gamma ^{\mu }\partial^{\nu }X^{8}\partial_{\mu }X^{i}\partial_{\nu }X^{i} \nonumber\\
& & { }+2\Gamma ^{\mu }\partial_{\mu }X^{8}\partial^{\nu }X^{i}\partial_{\nu }X^{i}+2\Gamma ^{\mu }\Gamma ^{i}\partial^{\nu }X^{8}\partial_{\nu }X^{8}\partial_{\mu }X^{i} \nonumber\\
& & { } -2 \Gamma ^{\mu }\Gamma ^{i}\partial^{\nu }X^{8}\partial_{\mu }X^{8}\partial_{\nu }X^{i}-2 \Gamma ^{\mu }\Gamma ^{i}\partial_{\mu }X^{j}\partial^{\rho  }X^{j}\partial_{\rho  }X^{i} \nonumber\\
& & { }-\Gamma ^{ij}\epsilon ^{\rho \sigma \lambda }\partial_{\rho  }X^{i}\partial_{\lambda  }X^{8}\partial_{\sigma  }X^{j})\epsilon  \nonumber\\
& & { }  + \lambda _{3}(-8 \Gamma ^{\mu \nu \rho }\Gamma ^{ijk}\partial_{\mu    }X^{i}\partial_{\nu    }X^{j}\partial_{\rho    }X^{k})\epsilon. 
\end{eqnarray}
The requirement that these terms should be expressible in an SO(8) invariant form places constraints on the coefficients
\begin{eqnarray}\label{aaw}
 \delta \psi & = & +4\lambda _{1}\Gamma _{\mu }\Gamma ^{I}\partial^{\nu }X^{J} \partial^{\mu }X^{J} \partial_{\nu }X^{I}\epsilon  -2\lambda _{2}\Gamma _{\mu }\Gamma ^{I}\partial^{\mu }X^{J} \partial_{\nu }X^{J} \partial^{\nu }X^{I} \epsilon  \nonumber\\
& & {} +(2\lambda _{2}-8\lambda _{1})(\Gamma _{\mu }\partial_{\nu }X^{j}\partial^{\nu }X^{j}\partial^{\mu }X^{8}+\Gamma _{\mu }\Gamma ^{j}\partial_{\nu }X^{8}\partial^{\nu }X^{8}\partial^{\mu }X^{j})\epsilon  \nonumber\\
& & {} -8 \lambda _{3}\Gamma ^{\mu \nu \rho }\Gamma ^{ijk}\partial_{\mu    }X^{i}\partial_{\nu    }X^{j}\partial_{\rho    }X^{k}\epsilon -\lambda _{2}\Gamma ^{ij}\epsilon ^{\rho \sigma \lambda }\partial_{\rho }X^{i}\partial_{\lambda  }X^{8}\partial_{\sigma  }X^{j}\epsilon. 
\end{eqnarray}
We see that in order for the second line to vanish we require $\lambda _{2}=+4\lambda _{1}$. A little algebra in the third line reveals that $\lambda _{2}=+24\lambda _{3}$. The absolute values are dependent on the $\alpha ^{'2}$ conventions in ten-dimensional Super Yang Mills. Setting $\lambda _{1}=+\frac{1}{32}$, we find $\lambda _{2}=+\frac{1}{8}$ and $\lambda _{3}=+\frac{1}{192}$ which is in perfect agreement with the ten-dimensional supersymmetry transformations derived in \cite{higher1, higher2, higher3}. Furthermore, we are now able to write down an expression for the truncated higher order BLG fermion transformation
\begin{eqnarray}\label{aax}
 \delta \psi & = & \; l^{3}_{p} ( +\frac{1}{8} \Gamma _{\mu } \Gamma ^{I} \partial_{\nu } X^{J} \partial^{\nu } X^{J} \partial^{\mu } X^{I} 
\;\; - \frac{1}{4} \Gamma _{\mu } \Gamma ^{I} \partial^{\mu } X^{J} \partial_{\nu } X^J \partial^{\nu } X^{I} \nonumber\\
& & {} - \frac{1}{24} \epsilon ^{\mu \nu \rho } \Gamma ^{IJK} \partial_{\mu } X^{I} \partial_{\nu } X^{J} \partial_{\rho } X^{K})\epsilon.  
\end{eqnarray}
Once the coefficients are fixed, it is a simple yet tedious task to apply the dimensional reduction and duality transformation to the  `non-abelian' terms and re-write them in an SO(8) invariant form \cite{Low1}. The final answer for the $l_p^3$ correction to the BLG fermion supersymmetry transformation is \footnote{This expression contains a few minor coefficient corrections compared to the result presented in \cite{Low1}.}
\begin{eqnarray}\label{aay}
 \delta \psi & = & \; l^{3}_{p} [ +\frac{1}{8} \Gamma _{\mu } \Gamma ^{I} D_{\nu } X^{J} D^{\nu } X^{J} D^{\mu } X^{I} 
\qquad \quad- \frac{1}{4} \Gamma _{\mu } \Gamma ^{I} D^{\mu } X^{J} D_{\nu } X^J D^{\nu } X^{I} \nonumber\\
& & {} - \frac{1}{24} \epsilon ^{\mu \nu \rho } \Gamma ^{IJK} D_{\mu } X^{I} D_{\nu } X^{J} D_{\rho } X^{K} 
\qquad+ \frac{1}{8} \Gamma ^{\mu \nu } \Gamma ^{I} D_{\mu } X^{J} D_{\nu } X^{K} X^{JKI} \nonumber\\
& & {} + \frac{1}{8}  \Gamma ^{IJK} D_{\mu } X^{L} D^{\mu } X^{J} X^{ILK}
\qquad \qquad- \frac{1}{48} \Gamma ^{IJK} D_{\mu } X^{L} D^{\mu } X^{L} X^{IJK} \nonumber\\
& & {} + \frac{1}{48} \Gamma ^{\mu \nu } \Gamma ^{IJKLM} D_{\mu } X^{I} D_{\nu } X^{J} X^{KLM} 
\quad \;+ \frac{1}{8} \Gamma _{\mu } \Gamma ^{J} D^{\mu } X^{K} X^{KLM} X^{LJM} \nonumber\\
& & {}+ \frac{1}{32} \Gamma _{\mu } \Gamma ^{IJKLM} D^{\mu } X^{M} X^{IJN} X^{KLN} 
\quad \; + \frac{1}{48} \Gamma _{\mu } \Gamma ^{J} D^{\mu } X^{J} X^{KLM} X^{KLM} \nonumber\\
& & {} -\frac{1}{48} \Gamma_{\mu } \Gamma ^{IJKLM} D^{\mu } X^{N} X^{IJM} X^{KLN} 
\quad \;-\frac{1}{576} \Gamma ^{IJKLMNP} X^{IJQ} X^{KLQ} X^{MNP}\nonumber\\
& & {}-\frac{1}{32} \Gamma ^{IJM} X^{IKN} X^{KLN} X^{LJM} 
\qquad \qquad -\frac{1}{144} \Gamma ^{IJM} X^{KLN} X^{KLN} X^{IJM}] \epsilon. 
\end{eqnarray}
 
It is worth emphasising that our result is true for the most general class of three-algebra theories. In the specific case of a Euclidean three-algbera, we recover the result calculated by Richmond in \cite{Richmond1}. In particular terms $9$, $11$ and $12$ in the above expression are shown to vanish and terms $13$ and $14$ are shown to combine. In the case of Euclidean three-algebra, Richmond showed that the symmetrised trace prescription leads to a number of identities (see \cite{Richmond1, Richmond:2011hs } for details). Let us consider term $11$ as an example
\begin{eqnarray}\label{aaz}
 \delta \psi _{11}&=&\Gamma_{\mu } \Gamma ^{IJKLM} D^{\mu } X^{N} X^{IJM} X^{KLN} \\ \nonumber
&=& \Gamma_{\mu } \Gamma ^{IJKLM} D^{\mu } X^{N} [X^{KJM} X^{ILN}+X^{IKM} X^{JLN}+X^{IJK} X^{MLN}] \\ \nonumber
 &=& -3\Gamma_{\mu } \Gamma ^{IJKLM} D^{\mu } X^{N} X^{IJM} X^{KLN} 
\end{eqnarray} 
and therefore this term must be zero. Likewise terms $9$ and $12$ can be shown to vanish as a result of the same identity. Next, we can rewrite term 13 as
\begin{eqnarray}
-\frac{1}{32}\Gamma ^{I} \Gamma ^{JM}X^{IKN}X^{KLN}X^{LJM}=+\frac{1}{96} \Gamma ^{IJM}X^{KLN}X^{KLN}X^{IJM}
\end{eqnarray}
where we made use of the identity $\Gamma ^{MN}X^{LMN} X^{JKL} X^{IJK} =+\frac{1}{3} \Gamma ^{MN}X^{IMN}X^{LJK} X^{LJK}$. This term will now combine with term 14 to give
\begin{eqnarray}\label{abf}
+\frac{1}{288} \Gamma ^{IJM}X^{KLN}X^{KLN}X^{IJM} \epsilon. 
\end{eqnarray}
 Thus in the special case of a Euclidean BLG theory we find the following expression for the four-derivative corrected fermion supersymmetry transformation
\begin{eqnarray}\label{abg}
 \delta \psi & = & \; l^{3}_{p} [ +\frac{1}{8} \Gamma _{\mu } \Gamma ^{I} D_{\nu } X^{J} D^{\nu } X^{J} D^{\mu } X^{I} 
\qquad\quad\; - \frac{1}{4} \Gamma _{\mu } \Gamma ^{I} D^{\mu } X^{J} D_{\nu } X^J D^{\nu } X^{I} \nonumber\\
& & {} - \frac{1}{24} \epsilon ^{\mu \nu \rho } \Gamma ^{IJK} D_{\mu } X^{I} D_{\nu } X^{J} D_{\rho } X^{K} 
\qquad+ \frac{1}{8} \Gamma ^{\mu \nu } \Gamma ^{I} D_{\mu } X^{J} D_{\nu } X^{K} X^{JKI} \nonumber\\
& & {} + \frac{1}{8}  \Gamma ^{IJK} D_{\mu } X^{L} D^{\mu } X^{J} X^{ILK}
\qquad\qquad \;- \frac{1}{48} \Gamma ^{IJK} D_{\mu } X^{L} D^{\mu } X^{L} X^{IJK} \nonumber\\
& & {} + \frac{1}{48} \Gamma ^{\mu \nu } \Gamma ^{IJKLM} D_{\mu } X^{I} D_{\nu } X^{J} X^{KLM} 
\quad \; + \frac{1}{8} \Gamma _{\mu } \Gamma ^{J} D^{\mu } X^{K} X^{KLM} X^{LJM} \nonumber\\
& & {}+ \frac{1}{48} \Gamma _{\mu } \Gamma ^{J} D^{\mu } X^{J} X^{KLM} X^{KLM} 
\qquad \quad \; +\frac{1}{288} \Gamma ^{IJM} X^{KLN} X^{KLN} X^{IJM}] \epsilon. 
\end{eqnarray}
This result is in complete agreement with the result derived by Richmond in \cite{Richmond1}. We are now in a position to utilise the method outlined in the previous section to determine an expression for the higher order supercurrent. The associated supercharge can then be used to generate the scalar field and gauge field supersymmetry transformations. By way of example, we will now show how this works for the first term appeaing in \eqref{abg}. We first construct the supercurrent
\begin{eqnarray}
\bar{\epsilon} J^{\sigma } &=&-\bar{\psi } \Gamma ^{\sigma } \delta \psi \\ \nonumber
 & =& -\frac{1}{8}(\bar{\psi } \Gamma ^{\sigma }  \Gamma^{\mu} \Gamma^{I} \epsilon) D_{\nu} X^{J} D^{\nu} X^{J} D_{\mu} X^{I} .
\end{eqnarray}
from which we can determine the supercharge. Next we generate the scalar supersymmetry transformation
\begin{eqnarray}\label{abi}
\delta  X^{I} &=& i \bar{\epsilon} [Q, X^I] \\ \nonumber
 &=& \; +\frac{1}{8} \; (i \bar{\epsilon} \; \Gamma^{\mu } \Gamma ^{0}  \Gamma^{J} \psi ) \;  \int d^{2} \sigma \; [ D_{\nu} X^{K} D^{\nu} X^{K} D_ {\mu }X^{J}, X^{I}] \\ \nonumber
 &=& \; -\frac{1}{8} \; (i \bar{\epsilon} \;  \Gamma^{J} \psi ) \; D_{\nu} X^{K} D^{\nu} X^{K} \int d^{2} \sigma \; [ \partial _{0} X^{J}, X^{I}]  \\ \nonumber
 &=&   -\frac{1}{8} \; (i \bar{\epsilon} \; \Gamma^{I} \psi ) \; D_{\nu} X^{J} D^{\nu} X^{J}.
\end{eqnarray}
Applying this simple calculational method to the terms appearing in \eqref{abg} leads to the following expression for the higher order BLG scalar supersymmetry transformation
\begin{eqnarray}\label{abl}
 \delta  X^{I} & = & \; -\frac{1}{8} \; (i \bar{\epsilon} \; \Gamma^{I} \psi ) \; D_{\nu} X^{J} D^{\nu} X^{J}  + \frac{1}{4} \; (i \bar{\epsilon} \; \Gamma^{J} \psi ) \; D_{\nu} X^{I} D^{\nu} X^{J} \nonumber\\
& & {}+ \frac{1}{8} \; (i \bar{\epsilon} \; \Gamma^{\mu  \nu } \; \Gamma^{IJK} \psi ) \; D_{\mu} X^{J} D_{\nu} X^{K} 
 -\frac{1}{4} \; (i \bar{\epsilon} \; \Gamma^{\mu }  \; \Gamma^{M} \psi ) \; D_{\mu} X^{J} X^{IJM} \nonumber\\
& & {} -\frac{1}{24} \; (i \bar{\epsilon} \; \Gamma^{\mu } \; \Gamma^{IJKLM} \psi ) \; D_{\mu} X^{J} X^{KLM} 
 + \frac{1}{8} \; (i \bar{\epsilon} \; \Gamma^{L} \psi ) \;  X^{JKI} X^{JKL}\nonumber\\
& & {} -\frac{1}{48} \; (i \bar{\epsilon} \; \Gamma^{I} \psi ) \;  X^{KLM} X^{KLM}. 
\end{eqnarray}
Again, these terms are in perfect agreement with the terms derived by Richmond in \cite{Richmond:2011hs}. We now turn our attention to the gauge field supersymmetry transformations. Each of the ten fermion terms appearing in \eqref{abg} gives rise to a supercharge term which is then used to generate the gauge field transformation. As an illustrative example, we will now show how this works for the first term appearing in \eqref{abg}. Our starting point is the supercharge constructed from the first term in \eqref{abg} which can then be used to generate the corresponding gauge field term
\begin{eqnarray}\label{abs}
\delta A_j  &=& +\frac{1}{8}  (i \bar{\epsilon} \; \Gamma^{\mu } \Gamma ^{0}  \Gamma^{I} \psi )   \int d^{2} \sigma  [ D_{\nu} X^{J} D^{\nu} X^{J} D_ {\mu }X^{I}, A_j ] \\ \nonumber
&=&  -\frac{1}{4} (i \bar{\epsilon} \Gamma^{\nu } \Gamma ^{0}  \Gamma^{J} \psi )  \int d^{2} \sigma [ A_{i},A_{j} ]  D^{i} X^{I} D_ {\nu }X^{J} X^{I} \\ \nonumber
& & { }\quad -\frac{1}{8} (i \bar{\epsilon}  \Gamma^{i } \Gamma ^{0}  \Gamma^{I} \psi ) \int d^{2} \sigma [ A_{i},A_{j} ]  D^{\nu} X^{J} D_ {\nu }X^{J} X^{I} \\ \nonumber
 &=& \; -\frac{1}{4} \; (i \bar{\epsilon} \; \Gamma^{\nu } \Gamma ^{0}  \Gamma^{J} \psi ) \epsilon_{ij}   D^{i} X^{I} D_ {\nu }X^{J} X^{I}  -\frac{1}{8}(i\bar{\epsilon } \Gamma ^{i }\Gamma ^{0 } \Gamma ^{I}\psi ) \epsilon _{ij}  D^{\nu} X^{J} D_ {\nu }X^{J} X^{I}. \nonumber
\end{eqnarray}
The next step is to sum over the contracted world-volume indices and to note that $\bar{\epsilon} \Gamma^0 = - \bar{\epsilon} \Gamma^{12}$ such that we can write
\begin{eqnarray}\label{aca}
 \delta A_{j }  &=& \; +\frac{1}{4} \; (i \bar{\epsilon} \;  \Gamma^{2 }   \Gamma^{J} \psi \epsilon_{ij})   D^{i} X^{I} D_ {1 }X^{J} X^{I}  -\frac{1}{4} \; (i \bar{\epsilon} \; \Gamma^{1 }  \Gamma^{J} \psi \epsilon_{ij})   D^{i} X^{I} D_ {2 }X^{J} X^{I}\nonumber\\
   & & {}+\frac{1}{4} \; (i \bar{\epsilon} \; \Gamma^{12 } \Gamma ^{0}  \Gamma^{J} \psi \epsilon_{ij})   D^{i} X^{I} D_ {0 }X^{J} X^{I} +\frac{1}{8}(i\bar{\epsilon } \Gamma _{j } \Gamma ^{I}\psi )\; D^{\nu} X^{J} D_ {\nu }X^{J} X^{I}.
 \end{eqnarray}
By explicitly setting $i=1$ and $j=2$ the above expression can be re-written as
  \begin{eqnarray}\label{acb}
 \; \delta A_{j }  &=& \; -\frac{1}{4} \; (i \bar{\epsilon} \;  \Gamma_{i }   \Gamma^{J} \psi)   D^{i} X^{I} D_ {j }X^{J} X^{I}  +\frac{1}{4} \; (i \bar{\epsilon} \; \Gamma_{j }  \Gamma^{J} \psi )   D^{i} X^{I} D_ {i }X^{J} X^{I}\nonumber\\
 & & {} +\frac{1}{4} \; (i \bar{\epsilon} \; \Gamma_{j \nu \rho  }  \Gamma^{J} \psi )   D^{\rho  } X^{I} D^ {\nu   }X^{J} X^{I} +\frac{1}{8}(i\bar{\epsilon } \Gamma _{j } \Gamma ^{I}\psi )\; D^{\nu} X^{J} D_ {\nu }X^{J} X^{I}.
  \end{eqnarray}
which represents the spatial components of the desired gauge field transformation. We now conjecture that the same relationship holds true for the time-like component of the gauge field transformation and therefore
  \begin{eqnarray}\label{acc}
 \; \delta A_{\mu  }  &=& \; -\frac{1}{4} \; (i \bar{\epsilon} \;  \Gamma^{\nu  }   \Gamma^{J} \psi)   D_{\nu } X^{I} D_ {\mu  }X^{J} X^{I}  +\frac{1}{4} \; (i \bar{\epsilon} \; \Gamma_{\mu  }  \Gamma^{J} \psi )   D^{\nu } X^{I} D_ {\nu  }X^{J} X^{I}\nonumber\\
 & & {} +\frac{1}{4} \; (i \bar{\epsilon} \; \Gamma_{\mu  \nu \rho  }  \Gamma^{J} \psi )   D^{\rho  } X^{I} D^ {\nu   }X^{J} X^{I} +\frac{1}{8} (i\bar{\epsilon } \; \Gamma _{\mu } \Gamma ^{I}\psi)  D^{\nu} X^{J} D_ {\nu }X^{J} X^{I}. 
  \end{eqnarray}
Applying this method to the remaining nine terms in \eqref{abg} results in the following simplified expression for the higher order BLG gauge field supersymmetry transformation
\begin{eqnarray}\label{abm}
\delta A_{\mu  }  &=& \; -\frac{1}{4} \; (i \bar{\epsilon} \; \Gamma^{\nu  }  \Gamma^{J} \psi )   D_{\nu } X^{I} D_ {\mu  }X^{J} X^{I}
\qquad\qquad+\frac{1}{2} \; (i \bar{\epsilon} \; \Gamma_{\mu  \nu \rho  }  \Gamma^{J} \psi )   D^{\nu } X^{J} D^ {\rho  }X^{I} X^{I} \nonumber\\ 
 & & {}\; -\frac{1}{8} \; (i \bar{\epsilon} \; \Gamma_{\mu }   \Gamma^{I} \psi ) \;  D_{\nu} X^{J} D_ {\nu }X^{J} X^{I}
\qquad\quad\; -\frac{1}{4} \; (i \bar{\epsilon} \;  \Gamma_{\mu  }   \Gamma^{J} \psi)   D_{\nu } X^{I} D^ {\nu  }X^{J} X^{I} \nonumber\\ 
 & & {}\; +\frac{1}{4} \; (i \bar{\epsilon} \; \Gamma^{\nu }   \Gamma^{J} \psi ) \;  D_{\mu} X^{I} D_ {\nu }X^{J} X^{I}
\qquad\quad\; +\frac{1}{4} \; (i \bar{\epsilon} \; \Gamma^{\nu }   \Gamma^{I} \psi ) \;  D_{\mu} X^{J} D_ {\nu }X^{J} X^{I}\nonumber\\ 
 & & {}\; +\frac{1}{8}\; (i\bar{\epsilon } \; \Gamma _{\mu \nu \rho  } \Gamma ^{IJK} \psi) \; D^{\nu }X^{J} D^{\rho }X^{K} X^{I}
\quad\; +\frac{1}{4}\; (i\bar{\epsilon } \;  \Gamma ^{I} \psi) \; D_{\mu }X^{K} X^{JKI} X^{J}\nonumber\\
& & {}\; +\frac{1}{8}\;(i\bar{\epsilon } \; \Gamma _{\mu \nu } \Gamma ^{JLM} \psi)\; D^{\nu }X^{J} X^{ILM} X^{I}
\quad\;\; -\frac{1}{48} \; (i \bar{\epsilon} \;\Gamma _{\mu } \Gamma^{J} \psi ) \; X^{KLM}X^{KLM}X^{J}.
\end{eqnarray}
These terms are in complete agreement with the gauge field supersymmetry terms derived by Richmond. Thus we have shown, at lowest non-trivial order in Fermions, that we are able to derive the full set of four-derivative corrected BLG supersymmetry transformations, in perfect agreement with the literature.  Following the original approach of Bagger and Lambert, one would like to show that this algebra closes on to equations of motion. Both the scalar field and gauge field transformations derived in this section have been shown to close on-shell \cite{Richmond1}. However, closure of the fermion supersymmetry transformation requires knowledge of the quadratic fermion terms in $\delta \psi$ and the cubic fermion terms in $\delta X^I$ and $\delta A_\mu$. It is therefore imperative that the structure of the quadratic fermion corrections are elucidated for $\delta \psi$. This will then allow us to construct the corresponding supercurrent and supercharge and generate the cubic fermion corrections to the gauge field and scalar field transformations. This will then allow for the closure of the higher order fermion transformation which will in turn reveal the correct higher order fermion equation of motion. In the next section we take the first step towards achieving this goal by deriving the quadratic fermion corrections to $\delta \psi$.
 
\section{Quadratic Fermion Transformations}
In this section we will apply our method to determine the non-trivial quadratic fermion terms of the fermion supersymmetry transformation. Our start point is the quadratic fermion terms appearing in the ten-dimensional Super Yang Mills Fermion supersymmetry transformation
\begin{eqnarray}\label{aec}
\delta  \psi_{(3)}  \;&=&  \; \alpha ^{'2}( \lambda_4 \bar{\psi} \Gamma^M D^N \psi F_{MN} \epsilon   +    \lambda_5 \bar{\psi } \Gamma ^{MNP} D_{M}\psi  F_{NP} \epsilon)
\end{eqnarray}
Performing the dimensional reduction and applying the dNS duality transformation results in the following expression
\begin{eqnarray}\label{beast} 
\delta \psi_{(3)} &=& \; +\lambda_4 \bar{\psi} \Gamma^{\mu \nu} D_\mu \psi D_\nu X^8 \epsilon - \lambda_4 \bar{\psi} \Gamma^\mu [X^j , \psi ] D_\mu X_j \epsilon \nonumber \\
& & {} \; - \lambda_4 \bar{\psi} \Gamma^j D^\mu \psi D_\mu X_j \epsilon + \lambda_4 \bar{\psi} \Gamma^i [X^j , \psi ] X_{ij} \epsilon \nonumber \\
& & {} \; -2 \lambda_5 \bar{\psi} D_\mu \psi D^\mu X^8 \epsilon + 2 \lambda_5 \bar{\psi} \Gamma^{\mu \nu} \Gamma^j D_\mu \psi D_\nu X_j \epsilon \nonumber \\
& & {} \; -\lambda_5 \bar{\psi} \Gamma^\mu \Gamma^{ij} D_\mu \psi X_{ij} \epsilon + 2 \lambda_5 \bar{\psi} \Gamma_\lambda \Gamma^i  [X_i , \psi ] D^\lambda X^8 \epsilon \nonumber \\
& & {} \; + 2 \lambda_5 \bar{\psi}\Gamma^\mu \Gamma^{ij}  [X_i , \psi ] D_\mu X_j \epsilon + \lambda_5 \bar{\psi} \Gamma^{ijk} [X^i , \psi ] X_{jk} \epsilon 
\end{eqnarray}
The next task is to re-write these terms in an $SO(8)$ invariant form. Furthermore, we will see that the requirement of SO(8) invariance will place a constraint on the coefficients $\lambda_4$ and $\lambda_5$. We begin by noticing that the first, third, fifth and sixth terms combine to form a single $SO(8)$ invariant term
\begin{eqnarray}
(2\lambda_5 - \lambda_4 ) \bar{\psi} \Gamma^I D_\mu \psi D^\mu X^I \epsilon
\end{eqnarray}
where we made use of $\Gamma^{\mu \nu} = -\Gamma^\nu \Gamma^\mu + \eta^{\mu \nu}$ and discarded terms proportional to the fermion equation of motion $\Gamma^\mu D_\mu \psi$ which will not appear at this order of $l_p^3$. Next we look at the second and eighth terms appearing in \eqref{beast}. A little thought reveals that these two terms, when combined, can be expressed as
\begin{eqnarray}
& &  (+\lambda_5 \bar{\psi} \Gamma^\mu \Gamma^K \Gamma^{IJ} [X^I , X^J , \psi ] D_\mu X^K + \lambda_5 \bar{\psi}  \Gamma^\mu \Gamma^{IJK} [X^I , X^J , \psi ]D_\mu X^K \nonumber \\
& & -2 \lambda_5 \bar{\psi} \Gamma^\mu [X^i , \psi ] D_\mu X^i - \lambda_4 \bar{\psi} \Gamma^\mu [X^i , \psi ] D_\mu X^i) \epsilon.
\end{eqnarray}
We see that $SO(8)$ invariance requires that the last two terms cancel and therefore the coefficients are constrained
\begin{equation}
\lambda_4 = - 2 \lambda_5.
\end{equation} 
The remaining terms appearing in \eqref{beast} can also be re-written in an SO(8) invariant form. The final result for the quadratic fermion term is
\begin{eqnarray}
 \delta \psi &=& +4 \lambda_5 \bar{\psi } \Gamma ^{I}D_{\mu }\psi D^{\mu  }X^{I}\epsilon\nonumber + \lambda_5 \bar{\psi }\Gamma ^{\mu }\Gamma ^{K}\Gamma ^{IJ}[X^{I},X^{J},\psi ]D_{\mu }X^{K} \epsilon \nonumber\\
& & { }+\frac{1}{3} \lambda_5 \bar{\psi } \Gamma ^{\mu }\Gamma ^{IJK}D_{\mu }\psi [X^{I},X^{J},X^{K}]\epsilon - \lambda_5 \bar{\psi }\Gamma ^{IJK}[X^{I},X^{L},\psi ][X^{J},X^{K},X^{L}]\epsilon \nonumber\\
  & &{}  - \lambda_5 \bar{\psi } \Gamma ^{I   } [X^{K},X^{J},\psi ][X^{I},X^{J},X^{K}]\epsilon. \label{bbb}
\end{eqnarray}
It is remarkable that our method has uniquely determined the quadratic fermion supersymmetry transformation in BLG theory whilst at the same time constraining the coefficients appearing in ten dimensional Super Yang-Mills theory. It should now, in principle, be possible use this expression to construct the corresponding supercurrent and supercharge and generate the cubic fermion corrections to the gauge field and scalar field transformations respectively. We hope to report on this result in a future publication.

\section{Conclusions}

In this short paper we have presented a new method for determining the four-derivative corrected supersymmetry transformations of BLG theory. For the special case of Euclidean three-algbera, we were able to reproduce the well known result of Richmond \cite{Richmond1}. Furthermore, we were able to apply our method to determine, for the first time, the quadratic fermion corrections to the higher order fermion supersymmetry transformation. What is perhaps surprising about our approach is that it only depends on knowledge of the ten-dimensional fermion supersymmetry transformation. We have also seen that the requirement of $SO(8)$ invariance in $(2+1)$ dimensions constrains the $\alpha'^2$ coefficients appearing in the ten-dimensional Super Yang-Mills Theory.

The wealth of riches that emerge from such a modest start point are suggestive of a deeper explanation. A few points are worth mentioning. Firstly, our method is contingent on our ability to construct the supersymmetry current from the relation $\bar{\epsilon}J^\mu = - \bar{\psi} \Gamma^\mu \delta \psi$. In other words, it appears that only knowledge of the fermion supersymmetry transformation is required. This may be related to the fact that the supersymmetry current, R-current and energy-momentum tensor live within the same supergravity supermultiplet. Since the R-current only depends on the fermion field it follows that the supervariation of the R-current, and therefore the supercurrent, will also only depend on the fermion supersymmetry transformation. Secondly, in order to generate the gauge field supersymmetry transformation from the supercharge we had to assume a certain Poisson structure for the spatial coordinates. It is not clear at this stage how we are to interpret this assumption and we hope to return to this issue in a future publication. Finally, it should be possible to use the quadratic fermion term to derive the cubic fermion corrections to the gauge field and scalar field. It would then be possible to close the fermion supersymmetry algebra and determine the higher order fermion equation of motion. The supervariaton of this would uniquely determine the higher order bosonic equations of motion. These could then be used to construct a maximally supersymmetric Lagrangian. This currently represents work in progress.

\section*{Acknowledgements}

The authors would like to thank Professor Neil Lambert for enlightening and insightful discussions. Andrew Low would like to thank the students of Wimbledon High School for providing intellectual stimulation and motivation.

\newpage
\appendix
\section{Conventions}
The supersymmetry transformation parameter $\epsilon$ and the fermion $\psi$ of the Bagger-Lambert theory belong to the ${\bf{8}_s}$ and ${\bf{8}_c}$ representations of the $SO(8)$ R-symmetry and are 32-component spinors satisfying
\begin{equation} 
\Gamma ^{\mu \nu \rho }\epsilon = +\epsilon ^{\mu \nu \rho }\epsilon,   \quad  \Gamma ^{\mu \nu \rho }\psi  = -\epsilon ^{\mu \nu \rho }\psi. \label{a1}
\end{equation}
We assume that $\epsilon_{012} = -\epsilon^{012}$ and thus it follows that
\begin{eqnarray}
\epsilon_{\mu \nu \lambda} \epsilon^{\mu \rho \sigma} &=& - (\delta^\rho_\nu \delta^\sigma_\lambda - \delta^\rho_\lambda \delta^\sigma_\nu ) \\
\epsilon_{\mu \nu \lambda} \epsilon^{\mu \nu \sigma} &=& - 2 \delta^\sigma_\lambda 
\end{eqnarray}
The following relations follow from the chirality constraint \eqref{a1}
\begin{eqnarray}
\Gamma ^{\mu \nu \rho }\epsilon &=&+\epsilon ^{\mu \nu \rho }\epsilon \\
\Gamma ^{\mu \nu \rho }\psi  &=&-\epsilon ^{\mu \nu \rho }\psi  \\
\epsilon ^{\mu \nu \rho }\Gamma _{\nu \rho }\epsilon &=&-2\Gamma ^{\mu }\epsilon  \\
\epsilon ^{\mu \nu \rho }\Gamma _{\rho }\epsilon &=&+\Gamma ^{\mu\nu  }\epsilon\\
\epsilon ^{\mu \nu \rho }\Gamma _{\nu \rho }\psi  &=&+2\Gamma ^{\mu }\psi   \\
 \epsilon  ^{\mu \nu \rho }\Gamma _{\rho }\psi  &=&-\Gamma ^{\mu\nu  }\psi 
\end{eqnarray}
Our Gamma matrix conventions are as follows
\begin{eqnarray}
\{\Gamma ^{\mu },\Gamma ^{\nu }\}&=&2\eta ^{\mu \nu } \\
\bar{\epsilon } _{1}\Gamma ^{a_{1}a_{2}...a_{n}}\epsilon  _{2}&=&(-1)^{n(n+1)/2}\;  \bar{\epsilon } _{2}\Gamma ^{a_{1}a_{2}...a_{n}}\epsilon _{1}\\
\Gamma ^{8}\epsilon &=&+\epsilon\\
\Gamma ^{8}\psi  &=&-\psi.
\end{eqnarray}
The three-bracket $X^{IJK}$ appearing in the duality-transformed supersymmetry transformation is defined as
\begin{equation}
X^{IJK} = g^I_{YM} [X^J, X^K ] + g^J_{YM} [X^K, X^I] + g^K_{YM} [X^I , X^J].  \label{njn}
\end{equation}
with
\begin{equation}
g^I_{YM} = (0, \ldots, g_{YM}), \quad I=1,2, \ldots, 8. \label{nkn}
\end{equation}
In deriving the quadratic fermion terms in Section 4 we made use of the following expressions which follow directly from \eqref{njn} and \eqref{nkn}. Note, we have suppressed factors of $g_{YM}$ in what follows
\begin{eqnarray}\label{aea}
  \bar{\psi } \Gamma ^{IJK}[X^{I},X^{L},\psi ][X^{J},X^{K},X^{L}]\epsilon &\rightarrow& -\bar{\psi }\Gamma ^{ijk} \;[X^{i},\psi ]  X^{jk}\epsilon \\
 \bar{\psi }\Gamma ^{I }[X^{K},X^{J},\psi ][X^{I},X^{J},X^{K}]\epsilon &\rightarrow&+2\bar{\psi }\Gamma ^{i}[X^{j},\psi ]X^{ij}\epsilon\\
 \bar{\psi } \Gamma ^{\mu }\Gamma ^{IJK}[X^{I},X^{J},\psi ]D_{\mu }X^{K}\epsilon  &\rightarrow&  -2\bar{\psi }\Gamma ^{\mu }\Gamma ^{jk}[X^{j},\psi ]  D_{\mu }X^{k}\epsilon \\
\bar{\psi }\Gamma ^{\mu }\Gamma ^{IJK}D_{\mu }\psi [X^{I},X^{J},X^{K}]  \epsilon  &\rightarrow& -3\bar{\psi }\Gamma ^{\mu }\Gamma ^{ij}D_{\mu }\psi  X^{ij}\epsilon\\
  \bar{\psi }\Gamma ^{\mu } \Gamma ^{I}[X^{I},X^{K},\psi ]D_{\mu }X^{K}\epsilon  &\rightarrow& -\bar{\psi }\Gamma ^{\mu }\Gamma ^{i}[X^{i},\psi ]D_{\mu }X^{8}\epsilon -\bar{\psi }\Gamma ^{\mu }[X^{i},\psi ]D_{\mu }X^{i}\epsilon  \\
 \bar{\psi} \Gamma ^{I}D_{\mu }\psi D^{\mu }X^{I}\epsilon &\rightarrow&-  \bar{\psi} D_{\mu }\psi D^{\mu }X^{8}\epsilon+\bar{\psi} \Gamma ^{i}D_{\mu }\psi D^{\mu }X^{i}\epsilon   \\ \nonumber
  \bar{\psi } \Gamma ^{\mu} \Gamma ^{K}\Gamma ^{IJ}[X^{I},X^{J},\psi ]D_{\mu }X^{K}\epsilon  &\rightarrow& +2\psi \Gamma ^{\mu }\Gamma ^{jk}[X^{j},\psi ]D_{\mu }X^{k} + 2\psi \Gamma ^{\mu }[X^{k},\psi ]D_{\mu }X^{k}\epsilon \nonumber \\
 & & { } +2\psi \Gamma ^{\mu }\Gamma^{j}[X^{j},\psi ]D_{\mu }X^{8}\epsilon 
\end{eqnarray}


\providecommand{\href}[2]{#2}\begingroup\raggedright
\endgroup

\end{document}